\documentclass[11pt]{article}
\usepackage{amsmath,amssymb}
\def\ben{\begin{equation}}
\def\een{\end{equation}}

 \let\m=\mu \let\n=\nu  \let\p=\pi

\let\C=\Chi

\def\nn{\nonumber} \def\bd{\begin{document}} \def\ed{\end{document}}
\def\ds{\documentstyle} \let\fr=\frac \let\bl=\bigl \let\br=\bigr
\let\Br=\Bigr \let\Bl=\Bigl
\let\bm=\bibitem
\let\na=\nabla
\let\pa=\partial \let\ov=\overline
\newcommand{\be}{\begin{equation}}
\newcommand{\ee}{\end{equation}}
\def\ba{\begin{array}}
\def\ea{\end{array}}
\def\ft#1#2{{\textstyle{\frac{\scriptstyle #1}{\scriptstyle #2} } }}
\def\fft#1#2{{\frac{#1}{#2}}}
\def\del{\partial}
\def\vp{\varphi}
\def\sst#1{{\scriptscriptstyle #1}}
\def\oneone{\rlap 1\mkern4mu{\rm l}}
\def\td{\tilde}
\def\wtd{\widetilde}
\def\ie{{\it i.e.\ }}
\def\dalemb#1#2{{\vbox{\hrule height .#2pt
        \hbox{\vrule width.#2pt height#1pt \kern#1pt
                \vrule width.#2pt}
        \hrule height.#2pt}}}
\def\square{\mathord{\dalemb{6.8}{7}\hbox{\hskip1pt}}}
\newcommand{\ho}[1]{$\, ^{#1}$}
\newcommand{\hoch}[1]{$\, ^{#1}$}
\newcommand{\bea}{\setlength\arraycolsep{2pt} \begin{eqnarray}}
\newcommand{\eea}{\end{eqnarray}}
\newcommand{\ra}{\rightarrow}
\newcommand{\lra}{\longrightarrow}
\newcommand{\Lra}{\Leftrightarrow}
\newcommand{\bp}{\tilde \beta^\prime}
\newcommand{\tr}{{\rm tr} }
\newcommand{\Tr}{{\rm Tr} }
\def\0{{\sst{(0)}}}
\def\1{{\sst{(1)}}}
\def\2{{\sst{(2)}}}
\def\3{{\sst{(3)}}}
\def\4{{\sst{(4)}}}
\def\5{{\sst{(5)}}}
\def\6{{\sst{(6)}}}
\def\7{{\sst{(7)}}}
\def\8{{\sst{(8)}}}
\def\m{{\sst{(m)}}}
\def\n{{\sst{(n)}}}
\def\cA{{{\cal A}}}
\def\cB{{{\cal B}}}
\def\cF{{{\cal F}}}
\def\cG{{{\cal G}}}
\def\cH{{{\cal H}}}
\def\tV{\widetilde V}
\def\tW{\widetilde W}
\def\tH{\widetilde H}
\def\tE{\widetilde E}
\def\tF{\widetilde F}
\def\tA{\widetilde A}
\def\im{{{\rm i}}}
\def\tY{{{\wtd Y}}}
\def\ep{{\epsilon}}
\def\vep{{\varepsilon}}
\def\bD{{{\bar D}}}
\def\R{{{\mathbb R}}}
\def\C{{{\mathbb C}}}
\def\H{{{\mathbb H}}}
\def\CP{{{\mathbb C}{\mathbb P}}}
\def\RP{{{\mathbb R}{\mathbb P}}}
\def\Z{{{\mathbb Z}}}
\def\bA{{{\mathbb A}}}
\def\bB{{{\mathbb B}}}
\def\bC{{{\mathbb C}}}
\def\bD{{{\mathbb D}}}
\def\bE{{{\mathbb E}}}
\def\bZ{{{\mathbb Z}}}
\def\Re{{{\frak{Re}}}}
\def\Im{{{\frak{Im}}}}
\def\cosec{{\,\hbox{cosec}\,}}
\def\Gm{{\Gamma_{\!\! -}}}
\def\Gp{{\Gamma_{\!\! +}}}
\def\stan{{standard }}
\def\nonstan{{supernumerary }}
\def\p{{\partial}}
\def\kdel#1{{\fft{\del}{\del#1}}}

\def\bog{{Bogomolny }}
\def\om{{\omega}}

\newcommand{\nnr}{\nonumber \\}
\newcommand{\pd}{\partial}
\newcommand{\ud}{\textrm{d}}
\newcommand{\dTH}{T^{\prime \, 0}_\textrm{H}}
\newcommand{\dOi}{\Omega^{\prime \, 0}_i}

\newcommand{\tamphys}{\it George P. and Cynthia W. Mitchell  Institute
for Fundamental Physics and Astronomy,\\
Texas A\&M University, College Station, TX 77843-4242, USA}

\newcommand{\auth}{C.N. Pope
}

\begin{document}

\begin{flushright}
\hfill{
MIFP-10-01\ \ \ \ \ \ \ \  }
 %\hfill{
%\bf hep-th/yymmnnn}
\end{flushright}

%\vspace{25pt}
\begin{center}
{\large {\bf Homogeneous Einstein Metrics on $SO(n)$ }}

\vspace{15pt}
\auth

\vspace{10pt}
\hoch{}{\tamphys}

%\hoch{\ddagger}{\it Interdisciplinary Center of Theoretical Studies,
%USTC, Hefei, Anhui 230026, PRC}

\vspace{10pt}

{\it  DAMTP, Centre for Mathematical Sciences,
 Cambridge University,\\  Wilberforce Road, Cambridge CB3 OWA, UK}

\vspace{10pt}

\vspace{30pt}

\underline{ABSTRACT}
\end{center}

   It is well known that every compact simple Lie group $G$ admits an
Einstein metric that is invariant under the independent left and right 
actions of $G$.  In addition to this bi-invariant metric, with $G\times G$
symmetry, it was shown by D'Atri and Ziller that every compact simple Lie
group except $SU(2)$ and $SO(3)$ admits at least one further homogeneous
Einstein metric, invariant under $G\times H$, where $H$ is some proper
subgroup of $G$.  In this paper we consider the Lie groups $G=SO(n)$
for arbitrary $n$, and provide an explicit construction of $(3k-4)$ 
inequivalent homogeneous Einstein metrics on $SO(2k)$, 
and $(3k-3)$ inequivalent homogeneous Einstein metrics on $SO(2k+1)$.

\vspace{15pt}

\thispagestyle{empty}

%\pagebreak
%\voffset=0pt
%\setcounter{page}{1}

%\tableofcontents

%\addtocontents{toc}{\protect\setcounter{tocdepth}{2}}

%%%%%%%%%%%%%%%%%%%%%%%%%%%%%%%%%%%%%%%%

\newpage

\section{Introduction}

   The Einstein equation $R_{\mu\nu}=\lambda\, g_{\mu\nu}$ places constraints
on a subset of the full Riemann curvature tensor.  Since in $d$ dimensions
the Riemann tensor has $\ft1{12} d^2(d^2-1)$ algebraically independent
components, while the Ricci tensor has $\ft12 d(d+1)$ algebraically
independent components, it follows that the Einstein equation impose less
and less of a constraint on the curvature as $d$ increases. In $d=3$ the
count of Riemann and Ricci tensor components is the same, and in fact
one can express the Riemann tensor algebraically in terms of the
Ricci tensor.  Dimension $d=4$
is the lowest in which the Ricci tensor has fewer components than
the Riemann tensor.

   In consequence of these considerations, one can expect that the richness
of solutions of the Einstein equations should increase with increasing
dimension.  In this paper we consider one aspect of this question,
by seeking homogeneous Einstein metrics on certain group manifolds.  

    It is well known that every compact simple Lie group $G$ admits a 
bi-invariant Einstein metric, \ie one that is invariant under $G_L\times G_R$,
the direct product of independent left-acting  and right-acting transitive
actions of the group $G$.  If we take $T^a$ to be the generators of the
Lie algebra of $G$, then if $g$ denotes a group element in $G$ we may
define left-invariant 1-forms $\sigma_a$ by
%%%%%
\be
g^{-1} dg = \sigma_a\, T^a\,.
\ee
%%%%%
The bi-invariant metric, of the form $\tr (g^{-1} dg)^2$ is, with a suitable
choice of basis for $T^a$, given by
%%%%%
\be
ds^2= c\, \sigma_a^2\,,
\ee
%%%%%
where $c$ is a constant.

   It has been shown by D'Atri and Ziller \cite{datzil} that every simple
compact Lie group except $SU(2)$ or $SO(3)$ admits at least one additional
homogeneous Einstein metric.   These metrics are still invariant under a
transitive $G$ action (left or right, according to convention choice; we
shall choose the case where the full $G_L$ is preserved).  However, the
D'Atri and Ziller Einstein metric is invariant only under a proper subgroup 
of the right-acting $G$.  

    The general left-invariant metric can be written as
%%%%%
\be
ds^2 = x_{ab}\, \sigma_a\, \sigma_b\,,\label{squashed}
\ee
%%%%%
where $x_{ab}$ is a constant symmetric matrix.  For the metric to be 
Riemannian, the eigenvalues of $x_{ab}$ must all be positive.  
The D'Atri-Ziller Einstein metric on $G$ falls into the class (\ref{squashed}),
for some specific choice of $x_{ab}\ne c\, \delta_{ab}$.

   It is known in particular cases that there may exist yet more homogeneous
Einstein
metrics on a given simple compact group $G$, 
over and above the bi-invariant and D'Atri-Ziller examples.  (See,
for example, \cite{bohker} for a review.)  As far as we are aware, the
largest number that have been found explicitly in any example are the
six inequivalent homogeneous Einstein metrics on the 14-dimensional
exceptional group $G_2$, obtained in \cite{gilupo}.

   In principle, the task of finding left-invariant Einstein metrics of
the form (\ref{squashed}) is a purely mechanical one.  The Ricci tensor can 
be calculated as an algebraic function of the squashing parameters $x_{ab}$,
and the Einstein equation then reduces to a system of coupled algebraic
equations for the $x_{ab}$.  The problem in practice is that if one 
considers the most general symmetric tensor $x_{ab}$, the equations
become too complicated to be tractable, for essentially any compact simple
Lie group larger than $SU(2)$.

   In the paper \cite{gilupo}, which constructed 4 inequivalent Einstein
metrics on $SO(5)$ and 6 on $G_2$, the problem was greatly simplified by
focusing on restricted metric ans\"atze, involving only a small number
of independent components among the $x_{ab}$.  The choice for these
non-vanishing $x_{ab}$ was dictated by requiring invariance under some
chosen subgroup of the Lie group $G$.  It turned out that this provided a
rather fertile ground in which examples of Einstein metrics could be found.
It does not necessarily yield all the homogeneous Einstein metrics on the
group, but it does provide a relatively simple way of finding some of them.

    An important issue when looking for Einstein metrics is to be able
to recognise whether an ostensibly 
 ``new'' metric is genuinely new, or whether it is
instead just a repetition of a previously-obtained example, possibly
disguised by a change of basis.  A very useful tool in this regard is
provided by calculating some dimensionless invariant quantity built from the 
metric and the curvature.  The simplest example is
%%%%%
\be
I_1= \lambda^{d/2}\, V\,,\label{I1}
\ee
%%%%%
where $R_{ab}=\lambda g_{ab}$, $V$ is the volume of the space, and $d$ is
the dimension of the Lie group $G$.  If $I_1$ takes different values for two
Einstein metrics on $G$ then the two metrics are definitely inequivalent.
If $I_1$ is the same for two Einstein metrics then they may be equivalent,
and in practice they typically are.  
The volume form is given by $\sqrt{\det( x_{ab})}\, 
\sigma_1\wedge \sigma_2\wedge
\cdots\wedge \sigma_n$, and we may for convenience just take $V$ to be given by
%%%%%
\be
V= \sqrt{\det(x_{ab})}\,,
\ee
%%%%%
since the integration over $\sigma_1\wedge \sigma_2\wedge
\cdots\wedge \sigma_n$ will just produce a universal multiplicative constant
factor.

   Another dimensionless invariant that is sometimes useful is
%%%%%
\be
I_2 = R^{abcd} R_{abcd}\, \lambda^{-2}\,.\label{I2}
\ee
%%%%%

   In the present paper, we shall look for homogeneous Einstein metrics
on the entire class of Lie groups $SO(n)$.  Following the methodology
of \cite{gilupo}, we shall make simple metric ans\"atze adapted to
subgroups of $SO(n)$; specifically, we consider
%%%%%
\be
SO(p)\times SO(q)\subset SO(n)\,,\qquad p+q=n\,.
\ee
%%%%%
A convenient way to formulate the problem is to introduce the left-invariant
1-forms $L_{AB}$ for $SO(n)$, where $L_{AB}=-L_{BA}$ and $1\le A\le n$, etc.
These satisfy the Cartan-Maurer equations
%%%%%
\be
dL_{AB}= L_{AC}\wedge L_{CB}\,.
\ee
%%%%%

   If we decompose the fundamental $SO(n)$ index as $A=(i,\tilde i)$,
where $1\le i\le p$ and $p+1\le \tilde i\le n$, then the following 
metric ansatz is invariant under the right action of $SO(p)\times SO(q)$
(as well as, of course, the left action of $SO(n)$):
%%%%%
\be
ds^2 = \ft12 x_1\, L_{ij}\, L_{ij} + \ft12 x_2 L_{\tilde i\tilde j}
L_{\tilde i \tilde j} + x_3 L_{i\tilde j} L_{i\tilde j}\,.
\label{3para}
\ee
%%%%%
In the obvious orthonormal frame $e^1 = \sqrt{x_1}\, L_{12}$, etc., the
components of the Ricci tensor in the $SO(p)$, $SO(q)$ and
$SO(p+q)/(SO(p)\times SO(q))$ subspaces can easily be shown to be given by
%%%%%
\bea
SO(p):&& \hbox{Ric} = \Big(\fft{p-2}{2 x_1} + \fft{q x_1}{2 x_3^2}\Big)
       \hbox{ Id}\,,\nn\\
SO(q):&& \hbox{Ric} = \Big(\fft{q-2}{2 x_2} + \fft{p x_2}{2 x_3^2}\Big)
\hbox{ Id}\,,\nn\\
SO(p+q)/(SO(p)\times SO(q)):&& \hbox{Ric}= 
 \Big(\fft{p+q-2}{x_3} - \fft{(p-1) x_1}{2 x_3^2} - 
          \fft{(q-1) x_2}{2 x_3^2}\Big) \hbox{ Id}\,,
\label{ricci}
\eea
%%%%%
where $\hbox{Id}$ denotes the identity metrix in each subspace.
Thus solving the Einstein equation $R_{ab}=\lambda g_{ab}$ amounts to
solving the three conditions 
%%%%%
\bea
\fft{p-2}{2 x_1} + \fft{q x_1}{2 x_3^2}&=&\lambda\,,\nn\\
\fft{q-2}{2 x_2} + \fft{p x_2}{2 x_3^2}
&=&\lambda\,,\nn\\
\fft{p+q-2}{x_3} - \fft{(p-1) x_1}{2 x_3^2} - \fft{(q-1) x_2}{2 x_3^2}&=&
\lambda\,.\label{3eqs}
\eea
%%%%%

   Since the ansatz is symmetrical under the exchange $(p,x_1)\leftrightarrow
(q,x_2)$, we may without loss of generality assume that $0\le q\le p$.  There
are then two special cases that arise, namely when $q=0$ and $q=1$, and then
all other cases with $q\ge 2$ follow a generic pattern.

\bigskip
\noindent
{\bf The case $q=0$, $p=n$:}

   In this case, only the $SO(p)$ subspace occurs, and the metric is
simply the bi-invariant one
%%%%%
\be
ds^2 = \ft12 x_1 L_{ij} L_{ij}\,.
\ee
%%%%%
This has
%%%%%
\be
\lambda= \fft{(p-2)}{2 x_1}\,.
\ee
%%%%%

\bigskip
\noindent
{\bf The case $q=1$, $p=n-1$:}

   In this case, only the $SO(p)$ and the $SO(p+q)/(SO(p)\times SO(1))$
subspaces occur, and the metric takes the form
%%%%%
\be
ds^2 = \ft12 x_1 L_{ij} L_{ij} + x_3 L_{in} L_{in}\,.
\ee
%%%%%
 The first and third equations in (\ref{3eqs}) then imply
either
%%%%%
\be
x_1=x_3\,,\qquad \lambda = \fft{(p-1)}{2 x_3}\,,\label{q11}
\ee
%%%%%
or else
%%%%%
\be
x_1= \fft{(p-2)}{p}\, x_3\,,\qquad \lambda= \fft{(p-1)(p+2)}{2p x_3}\,.
\label{q12}
\ee
%%%%%
The solution (\ref{q11}) is just a repetition of the bi-invariant metric on
$SO(n)$ but (\ref{q12}) is an inequivalent Einstein metric; it is invariant
under $SO(n)\times SO(n-1)$ but not under $SO(n)\times SO(n)$.

\bigskip
\noindent
{\bf The cases $q\ge2$, $p=n-q$:}

   In all these cases, the three subspaces in (\ref{ricci}) are all non-empty,
and the metric takes the form (\ref{3para}).
The three equations (\ref{3eqs}) then imply
%%%%%
\bea
\lambda &=& \fft{q y^2 + p-2}{2 y x_3}\,,\label{lamsol}\\
x_2&=& \fft{[2-p + 2(p+q-2) y - (p+q-1)y^2]\, x_3}{(q-1) y}\,,\label{x2sol}
\eea
%%%%%
and then either $y=1$ or 
%%%%%
\bea
0&=& (p+q-1)[p^2+(p+q)(q-1)] y^3 \nn\\
&&-
 [p(q-1)(4q-7)+ q(q-1)(q-3)+ 2p^2(3q-5)+ 3 p^3] y^2\nn\\
&& + 
(p-2)[p(5q-7)+ 2(q-1)^2 +3p^2] y - (p-2)^2(p+q-1) \,,\label{ycubic}
\eea
%%%%%
where we have defined 
%%%%%
\be
y= \fft{x_1}{x_3}\,.
\ee
%%%%%
The $y=1$ solution just gives the bi-invariant metric again.
In general, the cubic equation (\ref{ycubic}) is nontrivial, yielding
three inequivalent solutions, and hence three inequivalent Einstein
metrics for each choice of $p$ and $q$.  There are two classes of special cases
where the cubic factorises over the rationals into a product of a linear and
a quadratic polynomial:

\begin{itemize}

\item[(1)] $q=2$.  Equation (\ref{ycubic}) then implies that
%%%%%
\be
y=\fft{(p+1)(p-2)}{p^2+p+2}\,,\label{ylin}
\ee
%%%%%
or else $y=[p\pm\sqrt{p+2}]/(p+1)$.  These latter two solutions imply
that $x_2=0$, and hence the metric is degenerate, but (\ref{ylin}) gives
a non-trivial, and inequivalent, Einstein metric.

\item[(2)] $q=p\ge 3$.  Equation (\ref{ycubic}) has the solutions
%%%%%
\be
y= \fft{p-2}{3p-2}\,,\qquad \hbox{or}\qquad 
 y= \fft{2p(p-1)\pm\sqrt{4p^3-5p^2+2p}}{p(2p-1)}\,.
\ee
%%%%%
The two solutions involving the square root give equivalent Einstein 
metrics (with the roles of $x_1$ and $x_2$ exchanged), and so in total we
obtain two further inequivalent Einstein metrics in this $q=p$ case.

\end{itemize}

  There are also isolated examples, such as $(p,q)=$ (6,3)
and (10,6), where the cubic again factorises over the rationals into
a product of linear and quadratic polynomials.  In these cases, all three
roots give inequivalent Einstein metrics.
 
   After including all the inequivalent partitionings of $n=p+q$, we find
in summary that the construction described here provides the following
numbers of inequivalent Einstein metrics on $SO(n)$:
%%%%%
\bea
SO(2k):&& (3k-4) \hbox{ inequivalent Einstein metrics}\,,\qquad k\ge 2\nn\\
SO(2k+1):&& (3k-3) \hbox{ inequivalent Einstein metrics}\,,\qquad k\ge2\,.
\eea
%%%%%
For example, for $SO(10)$ we have the partitions $(p,q)=$ (10,0), (9,1), (8,2),
(7,3), (6,4) and (5,5), giving $1+1+1+3+3+2=11$ inequivalent Einstein metrics.
For $SO(11)$ we have the partitions $(p,q)=$ (11,0), (10,1), (9,2), (8,3),
(7,4) and (6,5), giving $1+1+1+3+3+3=12$ inequivalent Einstein metrics.

   For a given partition of $n=p+q$, the Einstein metrics on $SO(n)$
that we have constructed here have the symmetry 
$SO(n)\times SO(p)\times SO(q)$, where $SO(n)$ acts transitively on the left, 
while $SO(p)\times SO(q)$ acts (intransitively) on the right.  It is manifest,
therefore, that such metrics on $SO(n)$ with different partitions $n=p+q$
are inequivalent.  (This can easily be confirmed, case by case, by
comparing the values of the invariants $I_1$ or $I_2$, defined in (\ref{I1})
and (\ref{I2}).)  In cases such as $3\le q < p$, where three different solutions
for the squashing parameters arise for a given $p$ and $q$, the invariants
$I_1$ or $I_2$ can be seen to take different values for the three cases, 
and hence the three Einstein metrics are indeed inequivalent.  (The same
is true for the $q=p\ge 3$ case, where the two inequivalent solutions 
can be seen to have different values for $I_1$ or $I_2$.)
 
    It should be emphasised that this construction certainly does not in
general exhaust
all the possibilities for homogeneous Einstein metrics on $SO(n)$.  We have
focused only on metric ans\"atze adapted to the various $SO(p)\times SO(q)$
subgroups in our discussion.  Other subgroups can, and in some cases
certainly do, give rise to further possibilities.

   What we have established, in this paper, is that the number of 
homogeneous Einstein metrics on compact simple Lie groups can grow without
limit as the dimension of the group increases.  Specifically, we have 
exhibited $(3k-4)$ inequivalent homogeneous Einstein metrics on the group 
manifold $SO(2k)$, and $(3k-3)$ on the group manifold $SO(2k+1)$.

\section*{Acknowledgements}

%%%%%%%%%%%%%%%%%%%%%%%%%%%

This research has been supported in part by DOE Grant DE-FG03-95ER40917.

\end{document}